\documentclass[a4paper,twocolumn,amsmath,amssymb]{revtex4-1}

\usepackage{graphicx}
\usepackage{amsfonts}
\usepackage{units}

\newcommand{\Vector}[1]{\mathbf{#1}}
\newcommand{\dVector}[1]{\dot{\bf #1}}

\newcommand{\Matrix}[1]{\text{\boldmath$\sf #1$}}

\begin{document}

\title{Performance Assessment of Variational Integrators for Thermomechanical Problems 
\thanks{Financial support of the German Research Foundation (DFG) under grant KE-1667 is gratefully acknowledged.}}



\author{Dominik~Kern} \email{dominik.kern@mb.tu-chemnitz.de}
\affiliation{ Chair of Applied Mechanics and Dynamics,\\ TU Chemnitz, Germany}
 
\author{Ignacio~Romero} 
\affiliation{           ETSI Industriales, Technical University of Madrid, Spain\\
           IMDEA Materials, Getafe, Spain}

\author{Sergio~Conde~Mart\'in}
\author{Juan Carlos Garc\'ia-Orden}
\affiliation{          ETSI de Caminos, Canales y Puertos, Technical University of Madrid, Spain}

\date{\today}


\begin{abstract} Structure-preserving integrators are in the focus of ongoing research because of their
distinguished features of robustness and long time stability. In particular, their
formulation for coupled problems that include dissipative mechanisms is still
an active topic.  Conservative formulations, such as the thermo-elastic case without
heat conduction, fit well into a variational framework and have been solved
with variational integrators, whereas the inclusion of viscosity and
heat conduction are still under investigation. To encompass viscous forces and the classical
heat transfer (Fourier's law), an extension of Hamilton's principle is required.
In this contribution we derive variational integrators for thermo-viscoelastic systems 
with classical heat transfer. Their results are compared for two discrete model problems vs. Energy-Entropy-Momentum methods. 
Such comparisons allow to draw conclusions about their relative performance, weaknesses and strengths.

\end{abstract}

\maketitle

\section{Introduction} 
In mechanics, Variational Integrators (VI) and Energy-Entropy-Momentum (EEM) methods are
the current state of the art in structure preserving time integration.  Both of these families of
methods have been used for over two decades  in many applications and have consistently demonstrated
their ability to solve evolution equations in a robust and accurate manner.
Although completely different in their genesis and background, they pursue the same
goal of accurate structure preservation, and compete as they are applied to ever more general
theories.

The concept of VI was introduced by Cadzow \cite{cadzow1970discrete} in the
seventies and comprehensively developed by the group at Caltech and worked out many special cases
(see the overview \cite{marsden2001discrete}).
The basic idea is to start the discretization directly from the variational principle, thus skip the formulation of differential equations and lead to one-step maps, which are implicitly determined by algebraic equations.
Originally developed for conservative systems, their
extension to dissipative effects is currently being researched (c.f. the recent works
\cite{mata2011variational, kern2014variational}). 
Viscosity and heat transfer are such effects of technical relevance in order to account for damping and temperature effects.
The geometric consequence of dissipation is the loss of symplecticity, however the performance of VIs does not deteriorate. 
They are robust and by design at least second-order accurate.  Typical applications are in astronomy and space mission design, and increasingly in robotics, where VIs offer practical advantages for feedback control.

\smallskip
EEM methods were introduced by Romero \cite{romero2009thermodynamically} and utilize the discrete derivative operator \cite{gonzalez1996time}, 
initially developed for the Energy-Momen\-tum method (EM) \cite{simo1992discrete}, 
to build thermodynamically consistent algorithms from the {\em geometric} structure revealed by the GE\-NE\-RIC formalism \cite{ottinger2005beyond}. 
Such formalism enables a unified expression of the evolution equations of any isolated thermodynamic system to be generated from the addition of the reversible and irreversible parts, 
which are directly related to the gradient of the total energy and the total entropy of the system in terms of the state vector, respectively. 
Due to the key properties of the discrete derivative operator, the resulting methods are automatically second-order accurate, 
energy-preserving and entropy-producing by design. In addition, 
first order accurate staggered methods could also be formulated in terms of entropy in such a way that each step remains thermodynamically consistent, 
see also \cite{romero2009thermodynamically}. 
Within this approach, different thermomechanical systems have successfully been addressed in terms of the so called entropy-based formulation, 
i.e using entropy as thermodynamic state variable, such as discrete thermo-elasticity \cite{romero2009thermodynamically}, 
discrete thermo-viscoelasticity \cite{garcia2012energy} and continuos nonlinear thermoelasticity \cite{romero2010algorithms1,romero2010algorithms2} with heat transfer. 
Very recently, a tem\-per\-a\-ture-based formulation for discrete thermo-elasticity has been proposed in \cite{conde2015cnsns}, 
overcoming the problems associated to the use of the entropy and hence fully complementing the GENERIC-based approach.

\smallskip The purpose of this contribution is the assessment of the precision and robustness of
variational integrators, in comparison to EEM methods and focusing on their conservation
properties.  To this end, the paper is structured as follows. In section \ref{sec_problem} the
problem of thermomechanical systems is defined in general.  In section \ref{sec_time} the time
discretization is described, namely a VI and two EEM
schemes, namely the energy-entropy-momentum method in entropy and in temperature formulation.  In
section \ref{sec_examples} these integrators are compared with each other at two discrete systems, a
planar single and a spatial double pendulum. Section \ref{sec_summary} concludes this performance
assessments.


\section{Problem Definition}
\label{sec_problem}
\begin{figure}[tbh]
\centering
\includegraphics[width=0.3\textwidth]{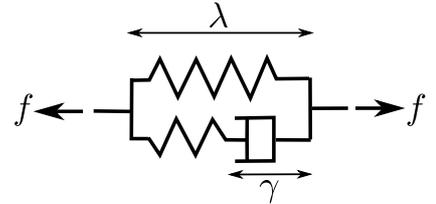} 
\caption{Generalized Maxwell element}
\label{fig_poynting}       
\end{figure}

In this article we consider thermo-viscoelastic models possessing a Lagrangian
\begin{equation}
L=T(\mathbf{q},\dot{\mathbf{q}})-\psi(\lambda, \gamma, \vartheta),
\end{equation}
where $T$ denotes kinetic coenergy 
(the distinction between kinetic energy and coenergy is in the spirit of Crandall
\cite{crandall1968dynamics}) and $\psi$, the free energy function (Helmholtz free energy). 
The structure of the free energy
\begin{equation} \label{eq_holzapfel}
\psi(\lambda, \gamma, \vartheta)=(1+\beta_c)\psi_e+\mu \gamma^2-\beta_c\gamma\frac{\partial \psi_e }{\partial \lambda}
\end{equation}
is adopted from Holzapfel and Simo \cite{holzapfel1996new}. 
This kind of models are referred to as generalized Maxwell-elements and may be represented as rheological model 
of a thermo-elastic spring (main spring), characterized by the free energy $\psi_e(\lambda,\vartheta)$, in parallel with 
a combination of another thermo-elastic spring in series with a dash-pot (see fig.~\ref{fig_poynting}). 
Sometimes it is also referred to as Poynting-element by some authors \cite{bertram2013festkoerpermechanik}. 
It may be used either as infinitesimal line element for the construction of a continuum model or 
as component of a discrete model. This element is described by three state variables, namely, the total stretch $\lambda$, the viscous stretch $\gamma$ and its temperature $\vartheta$.
For the temperature it will turn out useful to formulate it as time derivative $\vartheta=\dot{\alpha}$ of a quantity called thermacy $\alpha$, which is
also referred to as ``thermal displacement'', since temperature $\vartheta=\dot{\alpha}$ is related with
the averaged velocity of atoms (strictly speaking their averaged kinetic energy).
The corresponding momenta are the mechanical momentum $\mathbf{p}$ and the entropy $s$
\begin{subequations}
\begin{eqnarray}
\label{eq_pv}
 \mathbf{p}&=&\frac{\partial L}{\partial \dot{\mathbf{q}} }, \\
\label{eq_sT}
 s&=&\frac{\partial L}{\partial \dot{\alpha} }=-\frac{\partial \psi}{\partial \dot{\alpha}} .
\end{eqnarray}
\end{subequations}
Due to the similarities of these two definitions, 
the entropy is sometimes referred to as ``thermal momentum''.
The viscous stretch is related with a vanishing momentum variable, as its time derivative $\dot{\gamma}$ does not enter the Lagrangian.
The quantities conjugated to the deformation variables are total internal force $f$  and its viscous component $g$
\begin{subequations}
\begin{eqnarray}
 f&=&\frac{\partial \psi}{\partial \lambda},\\
 g&=&-\frac{\partial \psi}{\partial \gamma}.\label{eq_vf}
\end{eqnarray}
\end{subequations}
 The internal energy 
\begin{equation}
\label{eq_e}
e(\lambda, \gamma, \dot{\alpha})=\psi+\dot{\alpha}s
\end{equation}
is obtained by the Legendre transformation of the free energy $\psi$ with respect to the 
temperature $\dot{\alpha}$. Further relations that will be utilized later are
\begin{subequations}
\begin{eqnarray}
\label{eq_xt}
\dot{\mathbf{q}}&=&\frac{\partial \tilde{T}}{\partial \mathbf{p}}, \\
\label{eq_pt}
\dot{\alpha}&=& \frac{\partial \tilde{e}}{\partial s},
\end{eqnarray}
\end{subequations}
where $\tilde{T}(\mathbf{q,p})$ is the kinetic energy and 
$\tilde{e}(\lambda, \gamma, s)$ denotes the internal energy 
as a function of $(\lambda, \gamma, s)$.

\section{Time Discretization}
\label{sec_time}
The problems adressed in this paper are nonlinear in nature, and usually only numerical solutions can be obtained.
Three structure-preserving time integration schemes will be described in this section: a VI, an EEM method in entropy formulation (Gs) and in temperature formulation (GT).
All of them aim at good longtime behavior and improved numerical stability in comparison to standard methods for ordinary differential equations.

\subsection{Variational Integrators}
Variational integrators are based on a discrete version of Hamilton's principle
of stationary action. To formulate them, the essential step is thus to
write the action functional, from which the method follows naturally. For conservative
mechanical problems, this choice is standard. However, the inclusion of
dissipative effects (viscosity and Fourier's heat transfer) 
requires the use of incremental potentials \cite{yang2006variational} 
or the use of D'Alembert terms \cite{west2000variational}. In this article
we follow the latter option.

In this section the classical
Hamilton principle is extended to encompass thermomechanics. 
Based on this principle, variational integrators are constructed
in a systematic fashion. More specifically, the action is approximated
with a discrete action evaluated as a quadrature based on the midpoint rule (MP).
The resulting method, which we refer to as ``variational mid-point'', 
is not to be mistaken with the conventional
time integration scheme.  The generalization to higher order integrators will be discussed at the
end of this section.

\subsubsection{Variational Formulation}
The classical Hamilton principle reads
\begin{equation}
\delta S=\delta \int\limits_{t_0}^{t_1}L\,\mathrm{d}t=
\delta \int\limits_{t_0}^{t_1}(T-V)\,\mathrm{d}t=0, 
\end{equation}
where $L,T,V$ are, respectively, the Lagrangian, the kinetic and the potential energies.
The extension of this principle to thermoelasticity is obtained 
by replacing the potential energy $V(\mathbf{q})$ by the 
free energy $\psi(\lambda, \gamma, \vartheta)$ \cite{maugin2002hamiltonian}. 
The use of thermacy as variable makes the resulting Euler-Lagrange equations of the thermal problem to take the same mathematical structure as the mechanical
\begin{subequations}
\begin{eqnarray}
\frac{\mathrm{d}}{\mathrm{d} t}\frac{\partial L}{\partial \dot{\mathbf{q}}}-\frac{\partial L}{\partial \mathbf{q}}&=&0,\\
\frac{\mathrm{d}}{\mathrm{d} t}\frac{\partial L}{\partial \dot{\alpha}}-\frac{\partial L}{\partial \alpha}&=&0. 
\end{eqnarray}
\end{subequations}
These equations refer to the conservative case.
In order to include nonconservative generalized forces, Hamiltons principle is enhanced by D'Alembert terms \cite{west2000variational}
\begin{equation}
\delta \int\limits_{t_0}^{t_1}L\,\mathrm{d}t+\int\limits_{t_0}^{t_1}\delta W^\text{nc}\,\mathrm{d}t=0, 
\end{equation}
where $\delta W^\text{nc}$ splits into mechanical and thermal contributions
\begin{equation}
\delta W^\text{nc}_{\text{mech}}=\mathbf{f}\cdot\delta \mathbf{q}, \qquad  \delta W^\text{nc}_{\text{therm}}=\dot{s}\delta \alpha.  
\end{equation}
These terms account for mechanical forces such as external forcing or damping
and for entropy fluxes, also called ``thermal forces'' that stem from heat production (external source) or from heat transfer by Fourier's law.


\subsubsection{Construction of Variational Integrators}
The construction of a variational integrator starts directly from the variational principle. It consists of two steps.
Firstly, the functions of the generalized coordinates $\mathbf{q}(t)$ are discretized by interpolation functions $\mathbf{q}_d(t)$. 
Secondly, a quadrature rule for the evaluation of the action integral is applied. For the sake of simplicity the following explanations refer exemplarically to linear interpolation 
\begin{subequations}
\begin{eqnarray}
\mathbf{q}(t)&\approx&\mathbf{q}_d(t)=\frac{t^{k+1}-t}{t^{k+1}-t^k}\mathbf{q}^k+\frac{t-t^k}{t^{k+1}-t^k}\mathbf{q}^{k+1}, \\
\dot{\mathbf{q}}(t)&\approx&\dot{\mathbf{q}}_d(t)=\frac{\mathbf{q}^{k+1}-\mathbf{q}^k}{t^{k+1}-t^k} \ \ \text{for} \ \ \ t\in[t^k, t^{k+1}],
\end{eqnarray}
\end{subequations}
and numerical integration by the midpoint rule
\begin{equation}
\int\limits_{t^k}^{t^{k+1}}L\bigl(\mathbf{q}_d(t), \dot{\mathbf{q}}_d(t) \bigr)\,\mathrm{d}t
\approx \underbrace{h L \bigl(\mathbf{q}_d(t^{k+\nicefrac{1}{2}}), \dot{\mathbf{q}}_d(t^{k+\nicefrac{1}{2}}) \bigr)}_{L_d(\mathbf{q}^{k},\mathbf{q}^{k+1})},
\end{equation}
where $h=t^{k+1}-t^k$ denotes the time step size, $t^{k+\nicefrac{1}{2}}=t^k/2 + t^{k+1}/2$ half-time and $L_d$ is called discrete Lagrangian. 
As a result of this step, 
the continuous variational problem is turned into a discrete one, one requiring the stationarity
of the discrete action sum
\begin{equation}
S_d=(\mathbf{q}^0,\dots,\mathbf{q}^N)=\sum\limits_{k=0}^{N-1}L_d(\mathbf{q}^k,\mathbf{q}^{k+1}) 
\end{equation}
plus discrete D'Alembert terms 
\begin{subequations}
\begin{eqnarray}
\label{eq_stat}
\delta W^{nc}&=& \int\limits_{t_1}^{t_2} \mathbf{f}_d\cdot\delta \mathbf{q}  + \dot{s}\delta \alpha\,\mathrm{d}t \\
&\approx& \sum\limits_{k=0}^{N-1} h \bigl( \mathbf{f}\cdot\delta \mathbf{q}  + \dot{s}\delta \alpha  \bigr)|_{t=t_k+1/2}= \delta W^{nc}_d 
\end{eqnarray}
\end{subequations}
to vanish. The discrete forces, likewise the discrete Lagrangian, are obtained by numerical integration. If again
the midpoint rule is applied to the time integral, now of the virtual work, then the force during one time step
is split into discrete values, one at the beginning and the other at the end of the time step. Thus at each time point, except
the first and the last one, two forces enter the equation, one originating from the previous time step and the other from the next time step
\begin{subequations}
\begin{eqnarray}
\mathbf{f}_d^+ (\mathbf{q}^{k-1},\mathbf{q}^k)&=&\int\limits_{t^{k-1}}^{t^{k}}\mathbf{f}(t) \dfrac{\partial \mathbf{q}_d(t)}{\partial \mathbf{q}^k} \mathrm{d}t 
=\frac{h\mathbf{f} ( t^{k-\nicefrac{1}{2}} ) }{2},  \\ 
\mathbf{f}_d^- (\mathbf{q}^k,\mathbf{q}^{k+1})&=&\int\limits_{t^k}^{t^{k+1}}\mathbf{f}(t) \dfrac{\partial \mathbf{q}_d(t)}{\partial \mathbf{q}^{k}}  \mathrm{d}t 
=\frac{h \mathbf{f} ( t^{k+\nicefrac{1}{2}}  )}{2} ,
\end{eqnarray}
\end{subequations}
as deduced from the Discrete D'Alembert principle \cite{west2000variational}.
Evaluating the stationarity condition \eqref{eq_stat}
\begin{equation}
0= \sum_{k=0}^{N-1} \delta  L_d(\mathbf{q}^k,\mathbf{q}^{k+1})+\sum_{k=0}^{N-1}\delta W^{nc}_d(\mathbf{q}^k,\mathbf{q}^{k+1})
\end{equation} 
results in the $N$ discrete Euler-Lagrange equations for all admissable positions $\mathbf{q}^1$\dots$\mathbf{q}^{N}$.
The shorthand symbol $D_i$ will be used in the following to denote derivation with respect to the $i$th argument, i.e. $D_1L_d(\mathbf{q}^k,\mathbf{q}^{k+1})=\frac{\partial L_d}{\partial \mathbf{q}^k}$. 
The first equation
\begin{equation}
D_2L(\mathbf{q}^0, \dot{\mathbf{q}}^0)=-D_1L_d(\mathbf{q}^0,\mathbf{q}^{1}) - \mathbf{f}_d^- (\mathbf{q}^0,\mathbf{q}^{1}) \\
\end{equation}
determines $\mathbf{q}^{1}$ from the initial position $\mathbf{q}(t^0)=\mathbf{q}^0$ and initial velocity $\dot{\mathbf{q}}(t^0)=\dot{\mathbf{q}}^0$. 
While for $k=1\dots N-1$ the remaining positions  $\mathbf{q}^{k+1}$ follow recursively from
\begin{equation}
\begin{array}{rl}
0=& D_2L_d(\mathbf{q}^{k-1},\mathbf{q}^k)+\mathbf{f}_d^+ (\mathbf{q}^{k-1},\mathbf{q}^k)\\
& + D_1L_d(\mathbf{q}^k,\mathbf{q}^{k+1})+\mathbf{f}_d^- (\mathbf{q}^k,\mathbf{q}^{k+1}).
\end{array}
\end{equation}

Introducing the following definitions
\begin{subequations}
\begin{eqnarray} \label{eq_iter_lin}
 \mathbf{p}^k&=&-D_1L_d(\mathbf{q}^k,\mathbf{q}^{k+1})-\mathbf{f}_d^- (\mathbf{q}^k,\mathbf{q}^{k+1}),\\	%
 \mathbf{p}^{k+1}&=&+D_2L_d(\mathbf{q}^k,\mathbf{q}^{k+1})+\mathbf{f}_d^{+} (\mathbf{q}^k,\mathbf{q}^{k+1}),	%
\label{eq_update_lin}
\end{eqnarray}
\end{subequations}
results in the position-momentum form. This name is justified, because the $\mathbf{p}^k$ are the discrete analogues to momenta and fullfil a discrete version of Noether's theorem \cite{marsden2001discrete}.
Equation~\eqref{eq_iter_lin} is then solved iteratively with tangent matrix
\begin{equation}
\mathbf{T}=D_2D_1L_d(\mathbf{q}^k,\mathbf{q}^{k+1})-D_2\mathbf{f}_d^- (\mathbf{q}^k,\mathbf{q}^{k+1}) 
\end{equation}
by a Newton scheme. The resulting $\mathbf{q}^{k+1}$ are just inserted into
eq.~\eqref{eq_update_lin} in order to update the momentum. Then the procedure repeats for the next time step.

Higher order of the approximation increases the number of unknowns. For quadratic polynomials
\begin{subequations}
\begin{eqnarray}
q_d(t) &=&  a t^2+ b t+q^0 \qquad \text{with} \\
& &a = \frac{ 2q^0+2q^1-4q^{\nicefrac{1}{2}}}{h^{2}},\\
& &b = \frac{ 4q^{\nicefrac{1}{2}}-3q^0-q^1 }{h},
\end{eqnarray}
\end{subequations}
the unknows are $q^{\nicefrac{1}{2}}$ and $q^1$, since the value $q^0$ at the beginning of the time step is known.
A matching numerical integration is given by Simpson's rule
\begin{equation}
  L_d=\frac{h}{6} \left( L|_{t=t^k}+4L|_{t=t^{k+\nicefrac{1}{2}}}+L|_{t=t^{k+1}}\right).
\end{equation}
Evaluation of the variation results in the same number of Euler-Lagrange equations as the number of unknowns, here for the conservative part
\begin{subequations}
\begin{eqnarray}
\begin{array}{l}\boldsymbol{0}\\ \\ \end{array}&\begin{array}{l}=\\ \\ \end{array}&\begin{array}{l}
D_3 L_d(\mathbf{q}^{k-1},\mathbf{q}^{k-\nicefrac{1}{2}}, \mathbf{q}^k)\\
+D_1L_d(\mathbf{q}^k, \mathbf{q}^{k+\nicefrac{1}{2}},\mathbf{q}^{k+1}),                   
                 \end{array} \\
\boldsymbol{0}&=&D_2 L_d(\mathbf{q}^{k-1}, \mathbf{q}^{k-\nicefrac{1}{2}}, \mathbf{q}^k). 
\end{eqnarray}
\end{subequations}
Including nonconservative forces and expressing in position-momentum form yields the resulting equation system
\begin{subequations}
 \begin{eqnarray}
\label{eq_iter_1}
\mathbf{p}^k&=&-D_1L_d(\mathbf{q}^k, \mathbf{q}^{k+\nicefrac{1}{2}}, \mathbf{q}^{k+1})-\mathbf{f}_d^k, \\	
\label{eq_iter_2} 
\boldsymbol{0}&=&+D_2L_d(\mathbf{q}^k, \mathbf{q}^{k+\nicefrac{1}{2}}, \mathbf{q}^{k+1})+\mathbf{f}_d^{k+\nicefrac{1}{2}},\\ 	
\label{eq_update}
\mathbf{p}^{k+1}&=&+D_3L_d(\mathbf{q}^k, \mathbf{q}^{k+\nicefrac{1}{2}}, \mathbf{q}^{k+1})+\mathbf{f}_d^{k+1}, 	%
\end{eqnarray}
\end{subequations}
consisting of a nonlinear part, eqns.~\eqref{eq_iter_1}-\eqref{eq_iter_2}, which must be solved iteratively and 
an update equation \eqref{eq_update}.
While the first and last equation give the stepping from one time step to the next,
the second equation expresses the stationarity condition of the action during one time step.
Combining quadratic approximations with a Lobatto quadrature formula of third-order leads to a forth-order accurate VI scheme whereas the combination of linear approximation and mid-point rule is second-order accurate.

\smallskip
The state variables $\mathbf{q}$, $\mathbf{p}$ are calculated directly by the variational integrator. 
In contrast to mechanical position, thermacy is merely used for the derivation of the time stepping scheme, its absolute values are typically of little interest and hence not stored in simulations.
In order to obtain dependent quantitaties  $\dot{\mathbf{q}}$, i.e. velocity and temperature, their definitions by eqs.~\eqref{eq_xt}-\eqref{eq_pt} are rather evaluated than the time derivatives of their approximations.  
These values are better in terms of structure preservation than the interpolations, whose time-derivatives
are not necessarily continous at the time nodes. Similarly the kinetic energy $T$ and internal energy $e$ are evaluated as functions of the state variables $\mathbf{q}$, $\mathbf{p}$.
If these relations cannot be evaluated analytically, e.g. the Legendre transform does not provide an analytical expression
for the velocities as function of the momenta, they still can be evaluated numerically. 

\bigskip

One of the most important aspects of variational integrators is backward error analysis. It predicts the characteristics of the discrete-time path rather than the rate of convergence.
Considering the discretization of a Hamiltonian system by a variational integrator results in discrete states that are the exact solutions to a nearby Hamiltonian system \cite{hairer2006geometric}
\begin{equation}
 \tilde{H}(\mathbf{q},\mathbf{p})=H(\mathbf{q},\mathbf{p})+\frac{h^2}{2!}g_1(\mathbf{q},\mathbf{p})+\frac{h^4}{4!}g_2(\mathbf{q},\mathbf{p})+\dots,
\end{equation}
where $H$ denotes the original Hamiltonian. The functions $g_i$ can be determined using the method
of modified equations \cite{griffiths1986scope}.
Even though the conservation of the original Hamiltonian is incompatible with symplecticity, the energy error remains bounded \cite{hairer2006geometric}.
Variational integrators are symplectic as long as the time steps are equidistant.

Note that besides conservative forces variational integrators can handle both external forcing and dissipation \cite{marsden2001discrete} and proved well suited for practical applications in robotics \cite{johnson2009scalable} and also to be demonstrated by the following examples in this paper.

\subsection{Energy-Entropy-Momentum Methods}

For general, finite-dimensional isolated thermodynamic systems, the time-evolution of the state variables arranged in $\Vector{z}$, may be expressed by the following initial-value problem
\begin{equation}\label{eq:generic}
\dVector{z} = \Matrix{L}(\Vector{z})\nabla{E}(\Vector{z})+\Matrix{M}(\Vector{z})\nabla{S}(\Vector{z}), \quad \Vector{z}(0) = \Vector{z}_0\\
\end{equation}
$E$ being the total energy and $S$ being the total entropy, $\nabla (\bullet)$ being the gradient operator with respect to the state space vector,
  $\Vector{z}_0$ containing the prescribed initial conditions and $\Matrix{L}, \Matrix{M}$ being the so-called Poisson matrix and the Dissipative matrix, respectively. The evolution equations \eqref{eq:generic} will be in accordance with the laws of thermodynamics provided that the Poisson and Dissipative matrices are skew-symmetric and symmetric, positive semi-definite, respectively, and satisfy the degeneracy conditions
\begin{equation}\label{eq:degeneracy}
\nabla{S}^{\rm T}\Matrix{L}=\Vector{0}, \quad 
\nabla{E}^{\rm T}\Matrix{M}=\Vector{0}.
\end{equation}
The proof is straightforward and can be found in \cite{conde2015cnsns}. 

Following the guidelines in Romero \cite{romero2009thermodynamically}, the discrete derivative operator, denoted as $\mathsf{D}(\bullet)(\Vector{z}^{k+1}, \Vector{z}^k)$,
 is employed to arrive at the following monolithic implicit second order accurate method
\begin{equation}
\begin{split}
\frac{\Vector{z}^{k+1} - \Vector{z}^k}{h} &=  \Matrix{L}(\Vector{z}^{k+1}, \Vector{z}^k)\mathsf{D}E(\Vector{z}^{k+1}, \Vector{z}^k)\\
&+ \Matrix{M}(\Vector{z}^{k+1}, \Vector{z}^k)\mathsf{D}S(\Vector{z}^{k+1}, \Vector{z}^k)
\end{split}
\label{eq:discretegeneric}
\end{equation}
where the $\Matrix{L}(\Vector{z}^{k+1},\Vector{z}^k)$, $\Matrix{M}(\Vector{z}^{k+1},\Vector{z}^k)$ are second order approximations of the above Poisson and Dissipative matrices and,
 therefore, have their respective properties. Particularly, the degeneracy conditions are fulfilled in the following way
\begin{subequations}
\begin{eqnarray}
\label{eq:discretedegeneracya}
\mathbf{0}&=&\mathsf{D}S(\Vector{z}^{k+1}, \Vector{z}^k)^{\rm T}\Matrix{L}(\Vector{z}^{k+1}, \Vector{z}^k),\\
\mathbf{0}&=&\mathsf{D}E(\Vector{z}^{k+1}, \Vector{z}^k)^{\rm T}\Matrix{M}(\Vector{z}^{k+1}, \Vector{z}^k).
\label{eq:discretedegeneracyb}
\end{eqnarray}
\end{subequations}
The {\em discrete} laws of thermodynamics are thus satisfied due to these properties and the directionality property of the discrete derivative operator
 that allows to express the balance of any function in any time-step as
\begin{equation}
\begin{split}
E^{k+1} - E^k &= \mathsf{D}E\left(\Vector{z}^{k+1},\Vector{z}^k\right)\cdot(\Vector{z}^{k+1}-\Vector{z}^k)
\end{split}
\end{equation}
that can be further elaborated using \eqref{eq:discretegeneric} to give
\begin{equation}\label{eq:discrete1stlaw}
\begin{split}
&E^{k+1} - E^k\\
&=h\mathsf{D}E(\Vector{z}^{k+1}, \Vector{z}^k)^{\rm T}\Matrix{L}(\Vector{z}^{k+1}, \Vector{z}^k)\mathsf{D}E(\Vector{z}^{k+1}, \Vector{z}^k)\\
&+h\mathsf{D}E(\Vector{z}^{k+1}, \Vector{z}^k)^{\rm T}\Matrix{M}(\Vector{z}^{k+1}, \Vector{z}^k)\mathsf{D}S(\Vector{z}^{k+1}, \Vector{z}^k) = 0.
\end{split}
\end{equation}
Similarly, the total entropy balance results in
\begin{equation}\label{eq:discrete2ndlaw}
\begin{split}
&S^{k+1} - S^k\\
&= \mathsf{D}S\left(\Vector{z}^{k+1},\Vector{z}^k\right)\cdot(\Vector{z}^{k+1}-\Vector{z}^k)\\
&=h\mathsf{D}S(\Vector{z}^{k+1}, \Vector{z}^k)^{\rm T}\Matrix{L}(\Vector{z}^{k+1}, \Vector{z}^k)\mathsf{D}E(\Vector{z}^{k+1}, \Vector{z}^k)\\
&+
h\mathsf{D}S(\Vector{z}^{k+1}, \Vector{z}^k)^{\rm T}\Matrix{M}(\Vector{z}^{k+1}, \Vector{z}^k)\mathsf{D}S(\Vector{z}^{k+1}, \Vector{z}^k) \geq 0.
\end{split}
\end{equation}

The method \eqref{eq:discretegeneric} also ensures the conservation 
of quadratic momentum maps if the discrete derivative is modified to account for the
symmetries in the system. Details and proofs of this statement can be found in \cite{romero2009thermodynamically}. 

When specifying the presented approach, the Poisson and Dissipative matrices need to be fully defined. In doing so, 
a crucial issue in the formulation raises regarding the choice for the thermodynamic variables. Thus, in the very beginning of EEM methods the use of entropy variables was favored, as it easily provides the matrices and, therefore, thermodynamically consistent methods were straightforwardly achieved, although assuming significant restrictions in the formulation, such as difficulties for temperature boundary conditions. The recent work \cite{mielke2011formulation} concluded that a temp\-er\-a\-ture-based formulation can also provide the GENERIC matrices, facilitating the formulation of thermodynamically consistent methods based on temperature variables, see \cite{conde2015cnsns}, and thus overcoming the mentioned restrictions. In the following subsections we summarize the main aspects of both formulations for the model presented in Section \ref{sec_problem}. 

\subsubsection{Energy-Entropy-Momentum Methods in Entropy Formulation}

As previously pointed out, the GENERIC formulation only applies to isolated systems, i.e. the element and the environment
 with which it exchanges heat must be considered as the thermodynamic system. To this end, the easiest way is to consider the environment to have a constant temperature $\vartheta_\infty$. Thus the system is thermomechanically determined by means of five independent variables, among which both the element and the environment entropies $s$ and $s_\infty$ must be included to achieve an entropy-based formulation, that is 
\begin{equation}
\label{eq_G_state}
\Vector{z} = [\Vector{q},\Vector{p},\gamma,s,s_\infty]
\end{equation}

This choice simplifies the Poisson matrix to be the classical symplectic one, see \cite{Canonica1972}, and the Dissipative matrix gets not too involved, enabling such a straightforward formulation that it was achieved in \cite{garcia2012energy} with no need for the GENERIC form to reveal the structure meant to be preserved.  

However, this formulation is valid provided that the relations $\vartheta = \vartheta(\lambda,\gamma,s)$ could be analytically found. For standard temperature-based free energy functions \cite{dillon1962nonlinear}, this consideration limits the thermo-elastic parameters of the model to be at most linearly temperature-dependent. 

\subsubsection{Energy-Entropy-Momentum Methods in Temperature Formulation}
To avoid this issue, a temperature-based formulation becomes crucial, for which the element and the environment temperatures should be considered as state variables
\begin{equation}
\Vector{z} = [\Vector{q},\Vector{p},\gamma,\vartheta,\vartheta_\infty]
\end{equation}
Note that this choice implies the environment temperature $\vartheta_\infty$ to be non-constant so that the environment internal energy in terms of it, $\epsilon(\vartheta_\infty)\colon\mathbb{R^+}\rightarrow\mathbb{R}$, can be defined. With this consideration, the GENERIC matrices become more intricate but affordable (see \cite{conde2015cs}), thus allowing to formulate a temperature-based thermodynamically consistent counterpart which overcomes the restrictions related to the entropy formulation pointed out before. 


\section{Numerical Examples}
\label{sec_examples} 
The two classes of integrators introduced in the previous section, variational
integrators and EEM methods, are now to be applied to model problems from the
literature: a planar thermo-viscoelastic single pendulum \cite{garcia2012energy} and a spatial thermo-elastic
double pendulum \cite{conde2015cnsns}. The former compares a second-order accurate VI with a second-order accurate EEM method in entropy formulation whereas the latter compares the same VI with an EEM method in temperature formulation.

\subsection{Planar Thermo-Viscoelastic Single Pendulum with Classical Heat Conduction with the Environment} \label{sec_single}
\begin{figure}[tbh]
\centering
\includegraphics[width=0.175\textwidth]{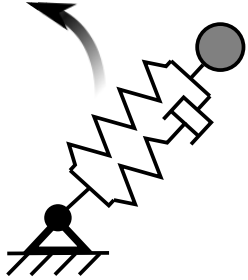} 
\caption{example 1: a mass attached to a massless thermo-viscoelastic spring with heat transfer with the environment}
\label{fig_sp_model}       
\end{figure}

This planar pendulum shown in Fig.~\ref{fig_sp_model} is taken from the literature \cite{garcia2012energy}. 
The length of the massless pendulum rod
\begin{equation}
  \lambda=\sqrt{x^2+y^2}
\end{equation} depends on the  position $(x,y)$ of the attached mass.
The evolution equation of the dash-pot is assumed to be linear
\begin{equation}
 \eta\dot{\gamma}=g, 
\end{equation}
where $\gamma$ denotes stretch of the dash-pot and $g$ the corresponding viscous force \eqref{eq_vf}.
The free energy, defined generally by eq.~\eqref{eq_holzapfel}, is specified by
thermo-elastic springs with 
\begin{equation} \label{eq_garcia}
\begin{array}{rl}
\psi_e(\lambda,\dot{\alpha})=&\dfrac{k}{2}\log^2\left(\frac{\lambda}{\lambda_0} \right)
-\beta_t(\dot{\alpha}-\vartheta_{r})\log\left(\frac{\lambda}{\lambda_0}\right)\\
 &+c \left[ \dot{\alpha}-\vartheta_r-\dot{\alpha}\log\left(\frac{\dot{\alpha}}{\vartheta_r}\right) \right],   
\end{array}
\end{equation}
where $k$ denotes the elasticity coefficient, which is related with stiffness, of the main spring, $\beta_t$ the thermomechanical coupling parameter, which is related with thermal expansion, and  $c$ the heat capacity.
This free energy function allows for large strains and Gough-Joule coupling.
Its parameters are summarized in tab.~\ref{tab_sp_para} for convenience.

The thermal part is the heat transfer between spring and environment and the heat generated by the dash-pot. The heat transfer is modeled by Fourier's law 
\begin{equation}
\phi=-\kappa (\dot{\alpha}-\vartheta_\infty),
\end{equation}
where the environment is assumed to be a thermal reservoir of constant temperature $\vartheta_\infty$.
Regarding the dash-pot, it is assumed that all energy mechanically dissipated is completely converted into heat, which corresponds to the entropy production
\begin{equation}
 \dot{s}_v=\frac{g\dot{\gamma}}{\dot{\alpha}}.
\end{equation}
The expression for the virtual work of the nonconservative forces thus consists of three summands
\begin{equation}
\delta W^{nc}
= -g\delta \gamma + \frac{g \dot{\gamma}}{\dot{\alpha}}  \delta\alpha
+\kappa \frac{\dot{\alpha}-\vartheta_\infty}{\dot{\alpha}} \delta \alpha
\end{equation}
 done by the viscous force of the dash-pot, heat generated by the dash-pot and heat conduction with the environment.
Both, energy and entropy of the environment are assumed zero initially. 

The free motion for given initial conditions is taken as example. The linearization around the
unstretched position at rest and at reference temperature indicates free oscillations of period
$t_\text{period}=0.086$s and gives an idea ot the motions time scale.

\smallskip
Using the same time step $h=0.2$s as in the reference \cite{garcia2012energy} results, after 3-4 Newton-iterations per time step, in trajectories and time histories of position and temperature that are indistinguishable. 

\begin{figure}[tbh]
\centering
\includegraphics[width=0.45\textwidth]{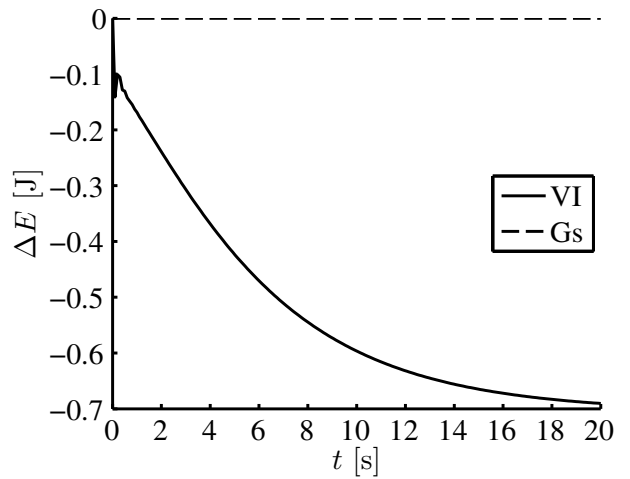}
\caption{example 1: error in total energy ($E=\text{const.}=1704J$) vs. time for the variational integrator (VI) and the EEM method (Gs)}
\label{fig_sp_energy}       
\end{figure}

\begin{figure}[tbh]
\centering
\includegraphics[width=0.45\textwidth]{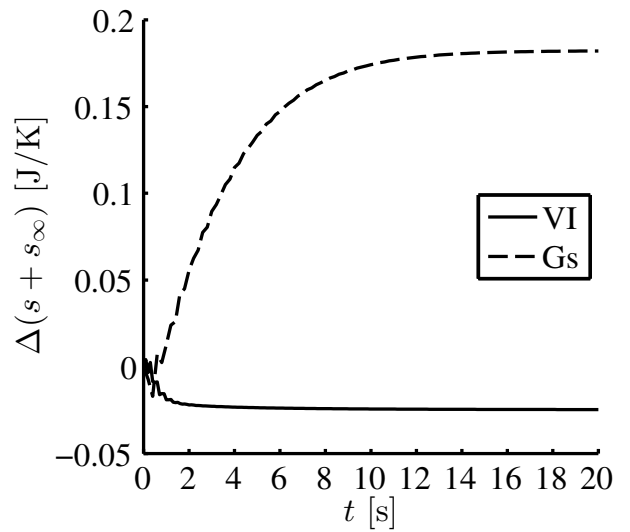}
\caption{example 1: error in total entropy ($s+s_\infty=5.4258\dots5.4613$J/K) vs. time for the variational integrator (VI) and the Energy-Entropy-Momentum method (Gs)}
\label{fig_sp_entropy}       
\end{figure}

Thus the focus is now on the conservation properties.
Fig.~\ref{fig_sp_energy}  shows how the numerically obtained total energy deviates from the exact value that is known to be constant on physical grounds.
As expected the EEM scheme outperforms the variational integrator in terms of energy as listed in tab.~\ref{tab_ex_error}.
The entropy is, as usual for diffusion processes in which temperatures level out, a monotonically increasing function asymptotically approaching its upper bound.
Fig.~\ref{fig_sp_entropy} shows the deviations of the total entropy compared to a reference trajectory much finer discretized than the other results ($h=0.005$s), 
both integrators show oscillatory deviations settling to the final value.
The entropy rates are not shown additionally, as they are in accordance with the second law,
i.e. strictly increasing for all the simulations, which is an inherent characteristic of both schemes.

Both integrators calculate the angular momentum, which is to be constant in this model, within machine precision.

Comparisons with standard solvers are not feasible since they get instable at this step size as documented in \cite{garcia2012energy}.

\begin{table}[t]
\caption{Single pendulum parameters (example 1) }
\centering
\label{tab_sp_para}       
\begin{tabular}{llll}
\hline\noalign{\smallskip}
$m$ & $1$&kg & mass\\
$\lambda_0$ & $1$&m & unstretched length\\
$k(\vartheta) $ & $k_{0}-k_{1}(\vartheta-\vartheta_r)$& & elasticity coefficient \\
$k_{0}$ & $100$&Nm\\
$k_{1}$ & $0.5$&Nm/K \\
$\beta_c$ & $0.1$& & spring ratio\\
$\beta_t$ & $4$&Nm/K & thermoelastic coupling\\
$c$ & $1$&Nm/K & heat capacity\\
$\mu(\vartheta) $ & $\mu_{0}-\mu_{1}(\vartheta-\vartheta_r)$& & viscosity coefficient\\
$\mu_0 $ & $5$&N/m \\
$\mu_1 $ & $0.1$&N/mK \\
$\eta(\vartheta) $ & $\eta_{0}e^{a(1/\vartheta-1/\vartheta_r)}$& & viscosity\\
$\eta_0$ & $100$&Ns/m \\
$a$ & $10$&K\\
$\kappa$ & $10$&W/K & thermal conductivity\\
$\vartheta_r$ & $300$&K & reference temperature\\
$\vartheta_\infty$ & $300$&K & environment \\
& & & temperature\\
$\mathbf{q}_0$ & $[3,0]$&m & initial position\\
$\dot{\mathbf{q}}_0$ & $[0,1]$&m/s & initial velocity  \\
$\gamma_0$ & $0$&m & initial viscous stretch\\
$\vartheta_0$ & $380$&K & initial temperature\\
& \\
$t_\text{sim}$ & $20$&s & simulation time\\
$h$ & $0.2$&s & time step\\
$\varepsilon$ & $10^{-10}$& & Newton tolerance for \\ 
& & & Euclidean norm $|\Delta \mathbf{z}|$ \\
\noalign{\smallskip}\hline
\end{tabular}
\end{table}

\subsection{Spatial Thermo-elastic Double Pendulum with Classical Heat Conduction}
\label{sec_double}
\begin{figure}[tbh]
\centering
\includegraphics[width=0.25\textwidth]{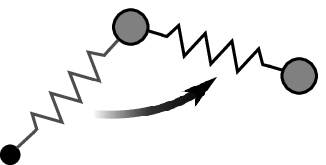} 
\caption{example 2: thermo-elastic double pendulum with heat transfer between the springs}
\label{fig_dp_m_model}      
\end{figure}
This example, shown in fig.~\ref{fig_dp_m_model}, is taken from Conde \cite{conde2015cnsns}. 
Its free energy is again given by eq.~\eqref{eq_garcia}, but its parameters, 
listed in tab.~\ref{tab_dp_para} are not only different in value but also in their functional dependencies. 
Particularly the logarithmic term in the elasticity coefficient turns the relation between entropy $s$ and temperature $\dot{\alpha}$ into a transcendental equation.
This makes the decisive difference compared to the previous example, the Legendre transform gets more involved, since there is no more analytical expression for temperature as
function of entropy. As a consequence, the EEM method in entropy formulation is ruled out.

The thermal system contains heat transfer between the springs but not with the environment. The heat fluxes from spring 1 into spring 2 and vice versa 
are again modeled by Fourier's law
\begin{subequations}
\begin{eqnarray}
\phi_1&=&-\kappa (\vartheta_1-\vartheta_2)=-\kappa(\dot{\alpha}_1-\dot{\alpha}_2),\\
\phi_2&=&-\phi_1.
\end{eqnarray}
\end{subequations}
Thus, the double pendulum forms an isolated system and its total energy $E=T+e$ is to be conserved.

The system linearized around the state of rest in vertical hanging position and at average temperature indicates a minimum period of free oscillations $t_\text{period}=0.17$s, which gives an orientation for setting the time step.

\begin{figure}[tbh]
\centering
  \includegraphics[width=0.45\textwidth]{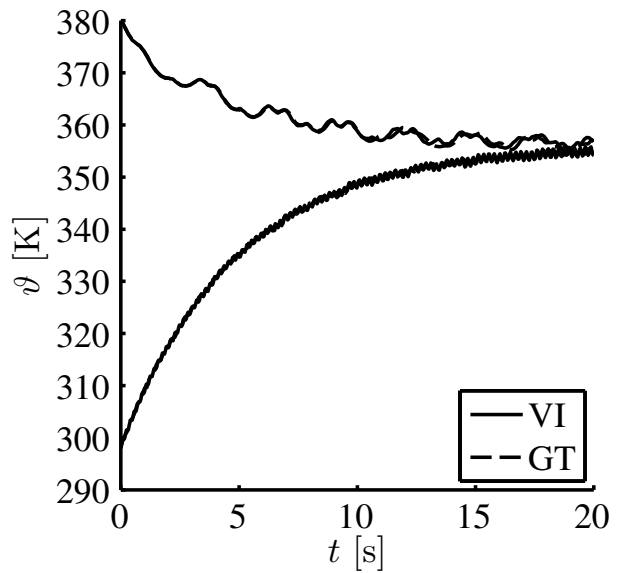}
\caption{Example 2: temperature vs. time for the variational integrator (VI) and the Energy-Entropy-Momentum method (GT)}
\label{fig_dp_tT}       
\end{figure}

\begin{figure}[tbh]
\centering
  \includegraphics[width=0.45\textwidth]{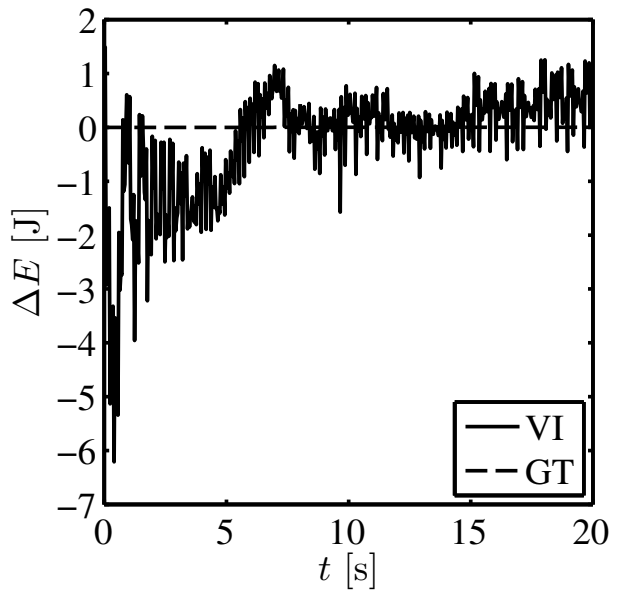}
\caption{Example 2: energy error ($E=\text{const.}=4\cdot10^5$J) vs. time for the variational integrator (VI) and the Energy-Entropy-Momentum method (GT)}
\label{fig_dp_energy}       
\end{figure}

Using the same time step $h=0.1$s  as Conde \cite{conde2015cnsns} makes the trajectories coincide in the beginning but diverge at about half-time ($t=12.5$s).
This divergence is probably rather due to the chaotic behavior of the mechanical system than to the integrator. The temperatures shown in 
fig.~\ref{fig_dp_tT} coincide well until the diverging mechanical behavior affects the temperatures by the relatively strong thermal coupling in this example.
Thus, the focus is more set on energetic quantities. 
Fig.~\ref{fig_dp_energy} shows total energy, which should be constant on physical grounds. 
As expected the EEM scheme outperforms the variational integrator. 

\begin{figure}[tbh]
\centering
  \includegraphics[width=0.45\textwidth]{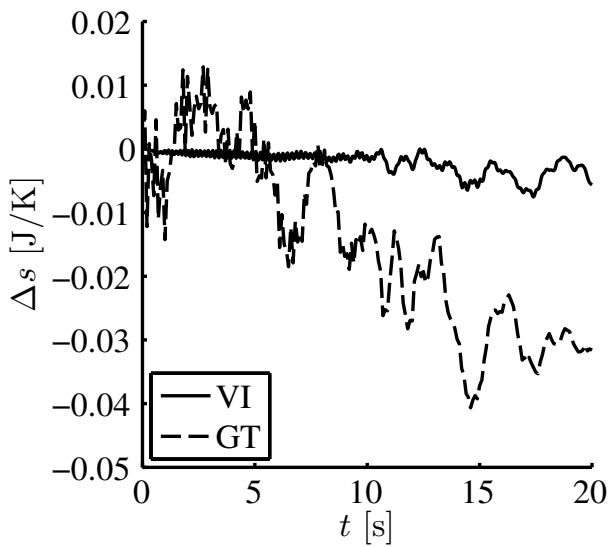}
\caption{Example 2: entropy ($s=1183\dots1222$J/K) error vs. time for the variational integrator (VI) and the Energy-Entropy-Momentum method (GT)}
\label{fig_dp_entropy}      
\end{figure}

The entropy in the discrete solution increases, as it should, in an isolated system.
A reference solution is taken to be a trajectory obtained with a very small
time step size ($h=0.005$s). Fig.~\ref{fig_dp_entropy} shows that the variational
integrator is closer to this reference trajectory than the EEM method.  Both integrators are in
accordance with the second law of thermodynamics by design.

\begin{figure}[tbh]
\centering
  \includegraphics[width=0.45\textwidth]{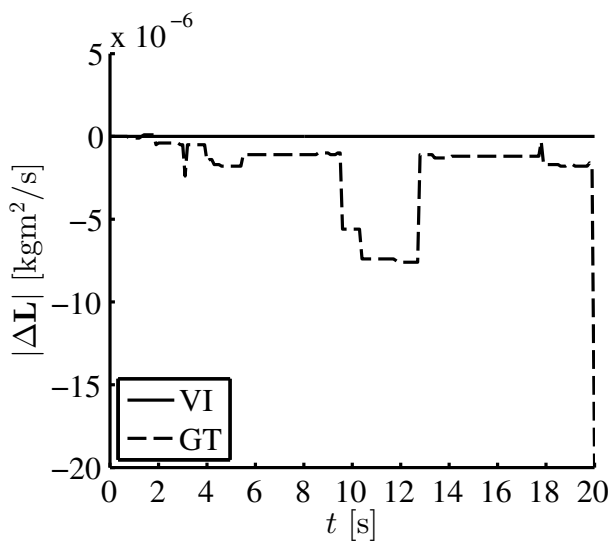}
\caption{Example 2: angular momentum ($|\mathbf{L}|=\text{const.}=71.6$kgm$^2$/s) vs. time for the variational integrator (VI) and the Energy-Entropy-Momentum method (GT)}
\label{fig_dp_momentum}      
\end{figure}

As the VI inherently preserves momentum maps it is supposed to perform better in the balance of
momentum.  For the double pendulum the angular momentum is to be preserved and indeed  in
fig.~\ref{fig_dp_momentum} only the error of the EEM scheme is visible while the variational
integrator stays within machine precision. In average, the variational integrator required one
Newton iteration (3-4 iterations) less then the EEM-integrator (4-5 iterations). We note that a standard solver (mid-point
rule with fixed step size) would have needed a time step size of less than $h=0.01$s in order to
stably integrate the motion. The relative errors for the simulations of both examples, double
pendulum and single pendulum simulations, are listed in tab.~\ref{tab_ex_error}. 

\begin{table}[t]
\caption{Double pendulum parameters (example 2)}
\centering
\label{tab_dp_para}       
\begin{tabular}{llll}
\hline\noalign{\smallskip}
$m_1$ & $10$&kg & mass\\
$m_2$ & $20$&kg & mass\\
$\lambda_{0,1}$ & $2$&m & unstretched length  \\
$\lambda_{0,2}$ & $1$&m & unstretched length \\
$k_i(\vartheta) $ & \multicolumn{2}{l}{$k_{i0}-k_{i1}\vartheta_r\log(\vartheta_i/\vartheta_r)$} & elasticity coefficient\\ 
$k_{10}$ & $5000$&J\\
$k_{11}$ & $50$&J/K \\
$k_{20}$ & $10000$&J\\
$k_{21}$ & $60$&J/K \\
$\beta_{t1}$ & $20$&J/K & thermoelastic coupling\\
$\beta_{t2}$ & $20$&J/K & thermoelastic coupling\\
$c_{1}$ & $5000$&J/K & heat capacity\\
$c_{2}$ & $2000$&J/K & heat capacity \\
$\kappa$ & $300$&W/K & thermal conductivity \\
$\vartheta_r$ & $300$&K & reference temperature\\
& \\
$\mathbf{q}_1(0)$ & $[3, 0, 0.5]^T$&m & initial position\\
$\mathbf{q}_2(0)$ & $[3, 1, 1]^T$&m & initial position\\
$\mathbf{p}_1(0)$ & $[0, 10, 0]^T$&kg m/s & initial momentum\\
$\mathbf{p}_2(0)$ & $[0, 0, -20]^T$&kg m/s  & initial momentum \\
$\vartheta_1(0)$ & $380$&K & initial temperature\\
$\vartheta_2(0)$ & $298$&K & initial temperature\\
& \\
$t_\text{sim}$ & $25$&s & simulation time\\
$h$ & $0.1$&s & time step\\ 
$\varepsilon$ & $10^{-10}$ & & Newton tolerance for\\
& & & Euclidean norm $|\Delta \mathbf{z}|$\\
\noalign{\smallskip}\hline
\end{tabular}
\end{table}

\begin{table}[t]
\caption{Relative errors in examples 1 and 2 computed as ratio between maximal deviation from the reference trajectory divided by the mean value of the reference trajectory}
\centering
\label{tab_ex_error}       
\begin{tabular}{llll}
\hline\noalign{\smallskip}
 & energy & entropy & momentum \\[3pt]
\hline
example 1: VI & $4\cdot 10^{-4}$ \% & $5\cdot10^{-4}$\% & $1\cdot 10^{-14}$\% \\
example 1: Gs & $3\cdot 10^{-15}$\% & $3\cdot10^{-3}$\% & $1\cdot 10^{-15}$\% \\
example 2: VI & $1\cdot 10^{-3}$ \% & $6\cdot10^{-4}$\% & $7\cdot 10^{-11}$\% \\
example 2: GT & $1\cdot 10^{-12}$\% & $3\cdot10^{-3}$\% & $3\cdot 10^{-5}$ \% \\
\hline
\end{tabular}
\end{table}


\section{Summary and Outlook}
\label{sec_summary} Variational integrators (VI) have been extensively employed for approximating the
evolution of Hamiltonian systems, leading to time integration schemes with remarkable, well-known,
features.  Similarly to Energy-En\-tro\-py-Mo\-men\-tum (EEM) methods, they  demonstrate that structure
preservation leads to time stepping schemes that are more robust than standard ones, especially in
stiff problems. Both time discretizations, VI and EEM methods, can employ
time step sizes which are too large for standard solvers, and their long term behavior is much more
accurate. In comparison to EEM methods, particularly in entropy formulation, the implementation of VIs poses no
restrictions on boundary conditions and parameter dependencies for the time stepping itself. However,
the postprocessing still may require further numerical evaluations. 

Thermo-viscoelasticity is not covered by classical Hamiltonian mechanics. However, we
have shown that viscous forces and heat transfer can be incorporated to the Hamiltonian action using
D'Alembert terms.
Comparison of the results for the discrete examples produced by same time steps shows the advantage of VI over EEM in momentum preservation, including entropy balance, on the one hand and the disadvantage of worse energy conservation on the other hand. This is not suprising, since EEMs are by design energy consistent and VIs known to preserve momentum maps exactly.
In addition it is observed for VI methods that they not get worse when applied to dissipative systems, i.e. when extended by the Discrete D'Alembert principle, where symplecticity is lost.

A general comment on both VI and EEM methods is that both, GENERIC-based and variational, demonstrate that structure preservation leads to algorithms more robust than standard ones.
As disadvantage they share the strong interlocking between physics and numerics making the implementation problem specific. 

Further works aims
at algorithmic speedup and simplification by separating and modularizing these integrators as much
as possible.
Yet another possible line of improvement is the development of splitting  methods that
separate the governing equations of the problem into implicit and explicit blocks. 
Incremental potentials \cite{yang2006variational} may be useful for this goal as
well as methods based on the discrete Pontryagin principle \cite{kharevych2006geometric}.



\bibliographystyle{spbasic}
\bibliography{compmech}   

%
%

\end{document}